\documentclass[pdflatex,sn-mathphys]{sn-jnl}

\jyear{2021}%

\theoremstyle{thmstyleone}%
\theoremstyle{thmstyletwo}%

\theoremstyle{thmstylethree}%

\usepackage{amsmath}
    {\end{bmatrix}\end{medsize}}%

\usepackage[left]{lineno}

\usepackage{etoolbox}
\makeatletter
\patchcmd{\ps@headings}
{\hbox to \hsize{\hfill Springer Nature 2021 \LaTeX\ template\hfill}}
{\hbox to \hsize{}}
{}
{}
\patchcmd{\ps@headings}
{\hbox to \hsize{\hfill Springer Nature 2021 \LaTeX\ template\hfill}}
{\hbox to \hsize{}}
{}
{}
\patchcmd{\ps@titlepage}
{\hbox to \hsize{\hfill Springer Nature 2021 \LaTeX\ template\hfill}}
{\hbox to \hsize{}}
{}
{}
\makeatother

\begin{document}

\title[~]{\vspace{-2cm}Ultrafast simultaneous manipulation of multiple ferroic orders through nonlinear phonon excitation}

\author[1]{\fnm{Daniel A.} \sur{Bustamante Lopez}}\email{dabl@bu.edu}
\equalcont{These authors contributed equally to this work.}

\author*[2]{\fnm{Dominik M.} \sur{Juraschek}}\email{djuraschek@tauex.tau.ac.il}
\equalcont{These authors contributed equally to this work.}

\author[3]{\fnm{Michael} \sur{Fechner}}\email{michael.fechner@mpsd.mpg.de}

\author[4]{\fnm{Xianghan} \sur{Xu}}\email{xx8060@princeton.edu}
\author[4]{\fnm{Sang-Wook} \sur{Cheong}}\email{sangc@physics.rutgers.edu}

\author*[1,5,6]{\fnm{Wanzheng} \sur{Hu}}\email{wanzheng@bu.edu}

\affil[1]{\orgdiv{Department of Physics}, \orgname{Boston University}, \orgaddress{\street{590 Commonwealth Avenue}, \city{Boston}, \postcode{02215}, \state{Massachusetts}, \country{USA}}}

\affil[2]{\orgdiv{School of Physics and Astronomy}, \orgname{Tel Aviv University}, \orgaddress{\postcode{6997801}, \state{Tel Aviv}, \country{Israel}}}

\affil[3]{\orgdiv{Max Planck Institute for the Structure and Dynamics of Matter}, \orgname{Center for Free-Electron Laser Science (CFEL)}, \orgaddress{\street{Luruper Chaussee 149}, \city{Hamburg}, \postcode{22761}, \country{Germany}}}

\affil[4]{\orgdiv{Rutgers Center for Emergent Materials and Department of Physics and Astronomy}, \orgname{Rutgers University}, \orgaddress{\street{136 Frelinghuysen Road}, \city{Piscataway}, \postcode{08854}, \state{New Jersey}, \country{USA}}}

\affil[5]{\orgdiv{Division of Materials Science and Engineering}, \orgname{Boston University}, \orgaddress{\street{590 Commonwealth Avenue}, \city{Boston}, \postcode{02215}, \state{Massachusetts}, \country{USA}}}

\affil[6]{\orgdiv{Photonics Center}, \orgname{Boston University}, \orgaddress{\street{8 Saint Mary's St.}, \city{Boston}, \postcode{02215}, \state{Massachusetts}, \country{USA}}}

\abstract{
Recent experimental studies have demonstrated the possibility of utilizing strong terahertz pulses to manipulate individual ferroic orders on pico- and femtosecond timescales. Here, we extend these findings and showcase the simultaneous manipulation of multiple ferroic orders in BiFeO$_3$, a material that is both ferroelectric and antiferromagnetic at room temperature. We find a concurrent enhancement of ferroelectric and antiferromagnetic second-harmonic generation (SHG) following the resonant excitation of a high-frequency fully-symmetric phonon mode. Based on first-principles calculations and phenomenological modeling, we ascribe this observation to the inherent coupling of the two ferroic orders to the nonequilibrium distortions induced in the crystal lattice by nonlinearly driven phonons. Our finding highlights the potential of nonlinear phononics as a technique for manipulating multiple ferroic order parameters at once. In addition, this approach provides a promising avenue to studying the dynamical magnetization and polarization behavior, as well as their intrinsic coupling, on ultrashort timescales. 
}




\maketitle


\section{Introduction}\label{sec1}
\begin{figure*}
\centering
\includegraphics[scale=0.075]{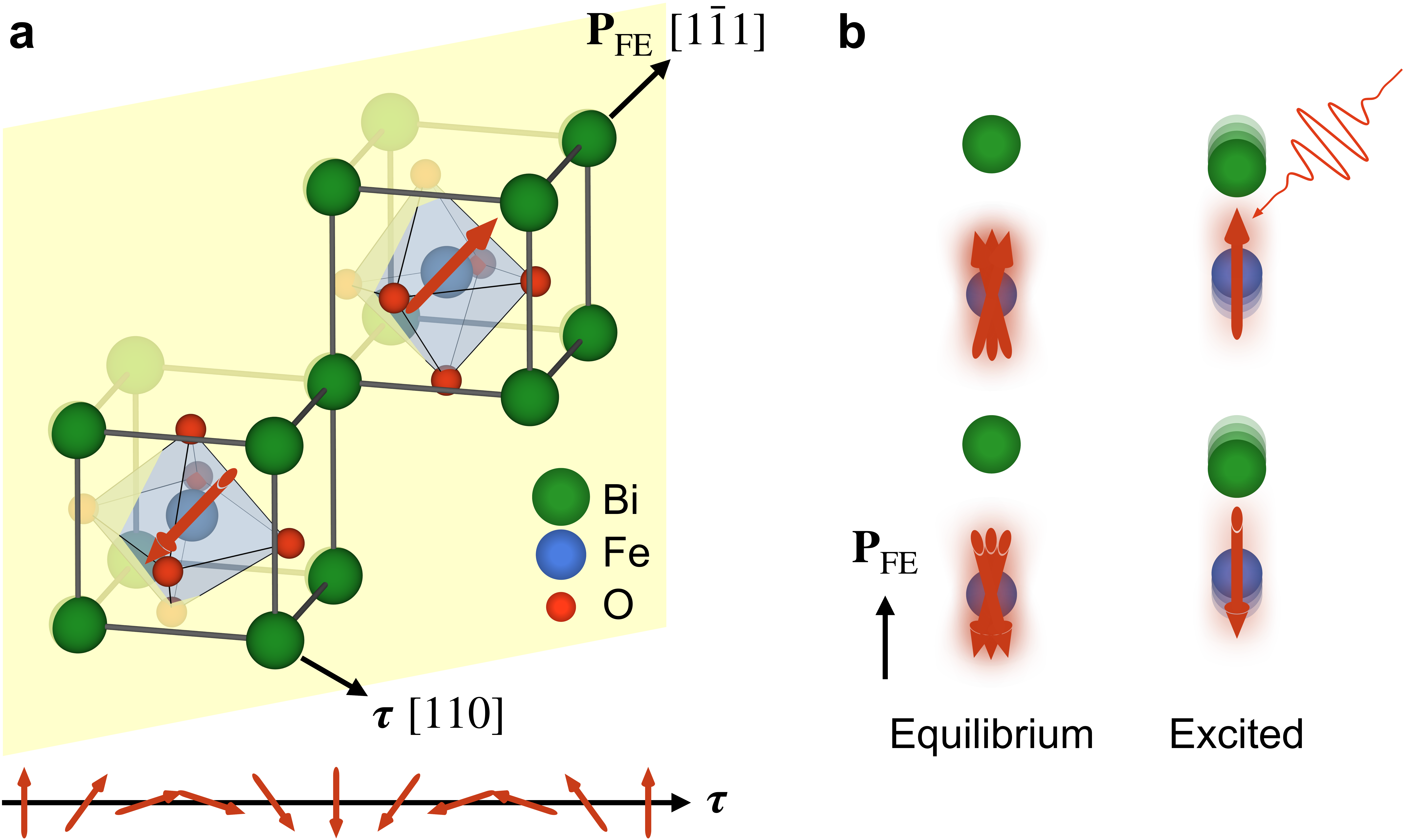}
\caption{\textbf{Control of multiferroic order through nonlinear phonon dynamics.}\newline{} \textbf{a,} Rhombohedral crystal structure of BiFeO$_3$ (space group $R3c$), in which the ferroelectric polarization ($\mathbf{P}_\mathrm{FE}$) is along the stretched body-diagonal of the pseudocubic unit cell, identified as $[1\bar{1}1]$ here. The spins between nearest-neighbor iron ions are antiferromagnetically aligned (red arrows) and form a spin cycloid with a wavelength of 62~nm along the normal vector $\boldsymbol{\tau}$ of the mirror plane (yellow) \cite{Lee2008, Ratcliff2016}. In a single-magnetic domain crystal, the spin cycloid breaks the three-fold rotation symmetry around $\mathbf{P}_\mathrm{FE}$. \textbf{b,} (Left) Initial state with the equilibrium ferroelectric polarization and a thermal broadening of the iron-spin alignment. (Right) The lattice distortion following the resonant excitation of a $A_1$ phonon leads to a change in polarization and to an improved alignment of the iron spins.
\label{fig:control}
}
\end{figure*}

Identifying efficient pathways to control the quantum phases of matter is a key challenge for the development of new materials with functional properties. In magnetoelectric multiferroics, the coupled magnetic and ferroelectric orders enable the control of magnetization using electric fields and vice versa, promising energy-efficient switching operations for data processing, and significant research efforts have been made to develop this concept within the last decade \cite{Eerenstein,SpaldinRamesh}. BiFeO$_3$, shown in Fig.~\ref{fig:control}a, is a well-known multiferroic material at room temperature, whose ferroelectric polarization can be switched by an applied electric field, which is accompanied by a reorientation of its antiferromagnetic domain \cite{ZhaoRamesh,LebeugleGukasov}. While this magnetoelectric coupling is appealing for low-energy consumption spintronics \cite{HeronRamesh2011,HeronRamesh2014}, the timescale of the switching process is limited by the ramp-up time of the applied electromagnetic fields, generally to nanosecond timescales.

In contrast, ultrashort laser pulses provide the possibility to generate large electric field strengths on sub-picosecond timescales, 
and recent experimental realizations of intense terahertz (THz) and mid-infrared (MIR) pulses with peak electric fields exceeding several megavolts per centimeter (MV~cm$^{-1}$) have opened up new pathways for dynamical materials engineering \cite{Liu2017,Vicario2020}. When the electric field component of a laser pulse is coupled resonantly to the dipole moment of an optical phonon mode, coherent lattice vibrations can be driven with amplitudes exceeding the harmonic regime, making anharmonic contributions of the interatomic potential-energy surface integral parts of the dynamics. A particular manifestation of this anharmonicitiy is nonlinear phononic rectification, a lattice analog to optical rectification, in which nonlinear coupling between phonon modes leads to a quasistatic distortion of the crystal structure \cite{Foerst2011,subedi:2014,fechner:2016,juraschek:2017}, creating transient crystal geometries that are not accessible in equilibrium. Examples of nonlinear phononic rectification have achieved to manipulate and induce ferroelectric \cite{subedi:2015,Mankowsky_2:2017,Nova2019,Li2019,Chen2022} and magnetic order \cite{Fechner2018,Khalsa2018,Rodriguez-Vega2020,Disa2020,Afanasiev2021,Stupakiewicz2021,Disa2021} individually. It has so far remained unexplored however, whether light can manipulate multiple ferroic order parameters simultaneously in a single domain.

Here, we coherently excite a high-frequency fully-symmetric ($A_1$) phonon mode in BiFeO$_3$ with intense MIR pulses and we probe the evolution of ferroelectricity and antiferromagnetism using time-resolved second-harmonic generation (SHG) at room temperature. We find that both the ferroelectric and antiferromagnetic contributions to the SHG signal are enhanced on a sub-picosecond timescale. We perform phonon dynamics simulations supported by density functional theory calculations that predict a rectification of the $A_1$ modes arising from both three-phonon coupling and cubic anharmonicities, whose polar atomic displacements increase the magnitude of the ferroelectric contribution to the second-order nonlinear susceptibility tensor. Comparison with experimental data confirms that the total change in SHG signal contains a significant proportion of antiferromagnetic contribution. Our calculations indicate that the rectified displacements along the $A_1$ modes lead to an enhancement of the magnetic interactions between the iron spins, which we interpret as a stabilization of the antiferromagnetic state (Fig.~\ref{fig:control}b).

\section{Results}\label{sec2}

\subsection{Phonon-driven enhancement of second-harmonic generation}\label{subsec2}

We study a single-crystal sample of BiFeO$_3$, cut and polished along the pseudocubic (110) plane. The ferroelectric polarization $\mathbf{P}_\mathrm{FE}$ is oriented along the body diagonal, [1$\bar{1}$1], which is parallel to the polished surface, as illustrated in Fig.~\ref{fig:control}a. Previous works show that the lattice, electronic, and magnetic degrees of freedom in BiFeO$_3$ are highly sensitive to external fields \cite{Lee2008, Choi2009,Kundys2010,Chen2015,Rubio-Marcos2018,Liou2019}. The large optical nonlinearities in BiFeO$_3$ \cite{Kumar2008} enables optical rectification that leads to coherent THz emission, which is a sensitive probe of the ferroelectric polarization \cite{Talbayev2008}. We observed a pronounced angular dependence of the emitted THz field amplitude that verifies that our sample is predominantly in a single domain (Supplementary Information Section S1).

\begin{figure*}
\centering
\includegraphics[width=1\textwidth]{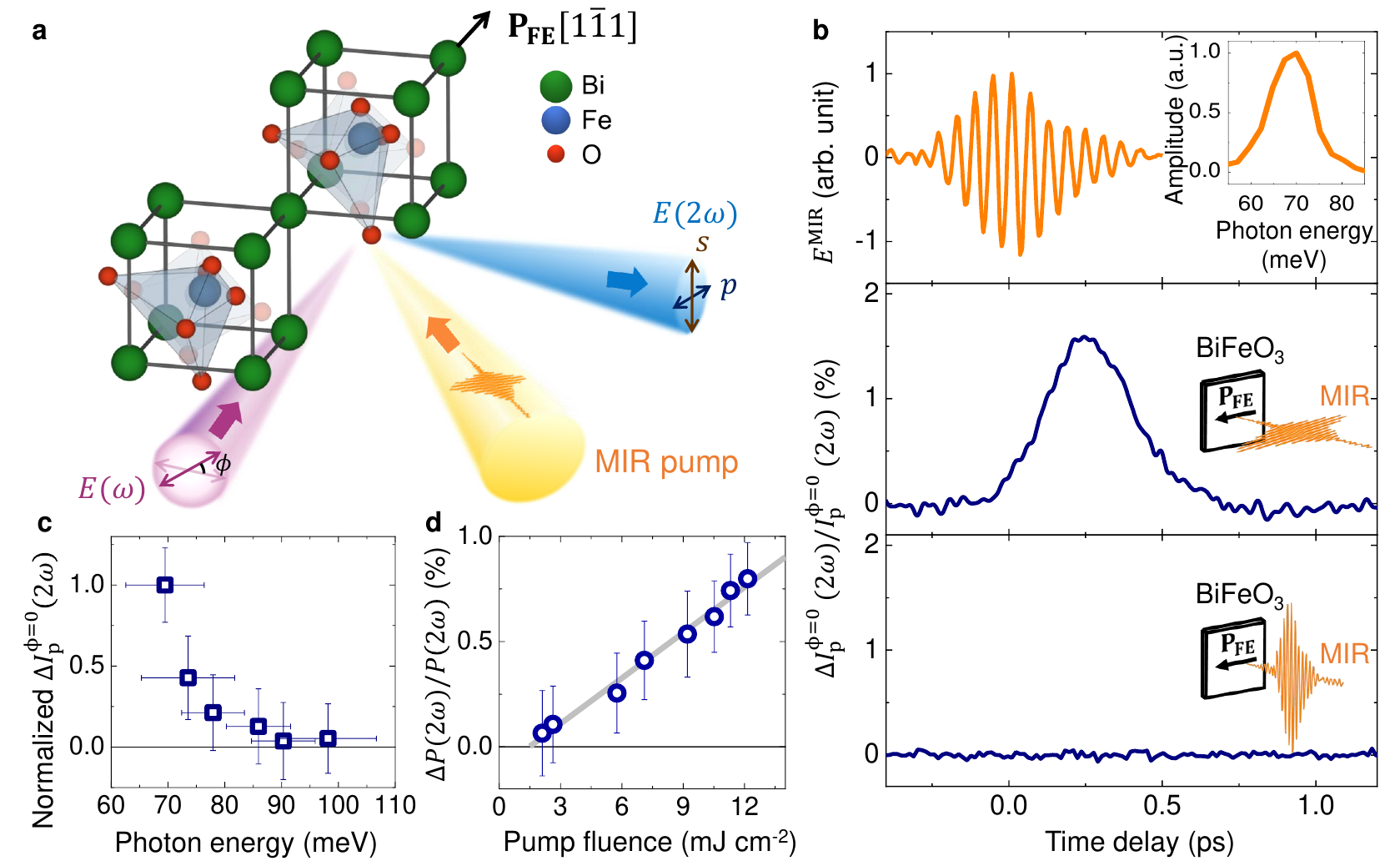}
\caption{\textbf{Phonon-driven enhancement of second-harmonic generation.} \textbf{a,} We used MIR pulses with a center frequency of 16.9~THz (69 meV) to drive the high-frequency $A_1$ phonon in BiFeO$_3$ at normal incidence. A probe beam at 800~nm was used for SHG with a 45$^{\circ}$ incidence in reflection geometry. \textbf{b,} (Top panel) Electric field component of the MIR pump and Fourier spectrum (inset). (Middle panel) The MIR pump induces a 1.5\% transient enhancement of the SHG intensity when polarized along $\mathbf{P}_\mathrm{FE}$. (Bottom panel) Rotating the MIR pump polarization by $90^{\circ}$ results in no pump-probe signal. \textbf{c,} Dependence of the SHG signal on the center frequency of the MIR pump pulse. The signal becomes nonzero as the frequency of the pump pulse approaches that of the high-frequency $A_1$ mode, indicating a phonon-driven mechanism. Vertical error bars represent the standard deviation. Horizontal bars are the bandwidths of the mid-infrared pump pulses. \textbf{d,} Linear dependence of the light-induced change in second-harmonic polarization on the pump fluence, consistent with nonlinear phononic rectification and ionic Raman scattering \cite{Neugebauer2021}. Vertical error bars represent the standard deviation. } 

\label{fig:MIRandDeltaISHG}
\end{figure*}

To control structural parameters critical to the ferroic properties, we used intense MIR pulses to excite the high-frequency $A_1$ mode in BiFeO$_3$ at 15.7~THz \cite{Palai2010}. We used frequency-tunable MIR pulses with an excitation fluence of 12 mJ~cm$^{-2}$ to couple to the electric dipole moment of the $A_1$ mode that is oriented along the ferroelectric polarization direction. Time-resolved SHG was used to probe the MIR-induced changes in the ferroic order parameters. The SHG intensity is proportional to the square of the light-induced nonlinear polarization, $I(2\omega)\propto\lvert P(2\omega) \rvert ^2$. The transient state is characterized by the light-induced change in SHG intensity, $\Delta I(2\omega)$, with the probe polarization oriented parallel to the light scattering plane (angle $\phi=0^{\circ}$) and the $p$-polarization detection configuration. A schematic of the pump-probe geometry is shown in Fig.~\ref{fig:MIRandDeltaISHG}a, indicating the atomic displacements induced by the high-frequency $A_1$ mode that are primarily composed of oxygen-ion motion. In Fig.~\ref{fig:MIRandDeltaISHG}b, we show the characterization of the MIR pump pulse and the polarization dependence of $\Delta I(2\omega)$. When the polarization of the MIR pulse is oriented along $\mathbf{P}_\mathrm{FE}$, the SHG intensity is enhanced by 1.5\%{}, whereas a polarization of the MIR pulse perpendicular to $\mathbf{P}_\mathrm{FE}$ yields no pump-induced change. When tuning the center frequency of the MIR pump pulse away from the $A_1$ phonon frequency, the pump-induced $\Delta I(2\omega)$ decays rapidly, as shown in Fig.~\ref{fig:MIRandDeltaISHG}c, indicating a resonant phonon-driven effect. Furthermore, the maximum transient change in polarization scales linearly with the pump fluence, shown in Fig.~\ref{fig:MIRandDeltaISHG}d, indicating a mechanism based on nonlinear phononic rectification and ionic Raman scattering \cite{Neugebauer2021}.


SHG is a sensitive probe for ferroelectricity and magnetic ordering, as the symmetry of the material is reflected by the coefficients of the SHG susceptibility tensor that is given by

\begin{equation}
\mathbf{P}(2\omega)\propto \left(\chi^{(i)}+\chi^{(c)} \right) : \mathbf{E}(\omega) \otimes \mathbf{E}(\omega),
\end{equation}
where $\mathbf{E}(\omega)$ is the electric field component of the laser pulse and $\chi^{(i)}$ and $\chi^{(c)}$ respectively are the ferroelectric and magnetic contributions to the SHG susceptibility tensor \cite{Denev2011, Fiebig2005}. The nonzero elements of the SHG tensors are determined by the point-group symmetry of the crystal. At room temperature, the crystal lattice of BiFeO$_3$ is rhombohedral with a $3m$ point-group symmetry (space group $R3c$). In the bulk, the antiferromagnetically aligned spins exhibit an additional long-range cycloidal modulation (Fig.~\ref{fig:control}a) that can be oriented along three equivalent wave vectors, commonly denoted by $(\boldsymbol{\tau}_1,\boldsymbol{\tau}_2,\boldsymbol{\tau}_3)$. The wave vectors $\boldsymbol{\tau}_i$ are perpendicular to $\mathbf{P}_\mathrm{FE}$ \cite{Ratcliff2016,Lee2008}.Our sample is predominantly in a single ferroelectric domain, as well as a single spin-cycloid domain, as discussed in Supplementary Information S1 and S3. The uniqueness of the spin-cycloid wave vector lifts the three-fold rotation symmetry around $\mathbf{P}_\mathrm{FE}$, and effectively lowers the symmetry of the system to monoclinic, described by point group $m$. Because the ferroelectric and magnetic properties determine the point-group symmetry, information about them is intrinsically encoded in the symmetry and magnitude of the SHG tensor components that can be probed by varying the polarization of the probe beam. 


\begin{figure*}
\centering
\includegraphics[width=1\textwidth]{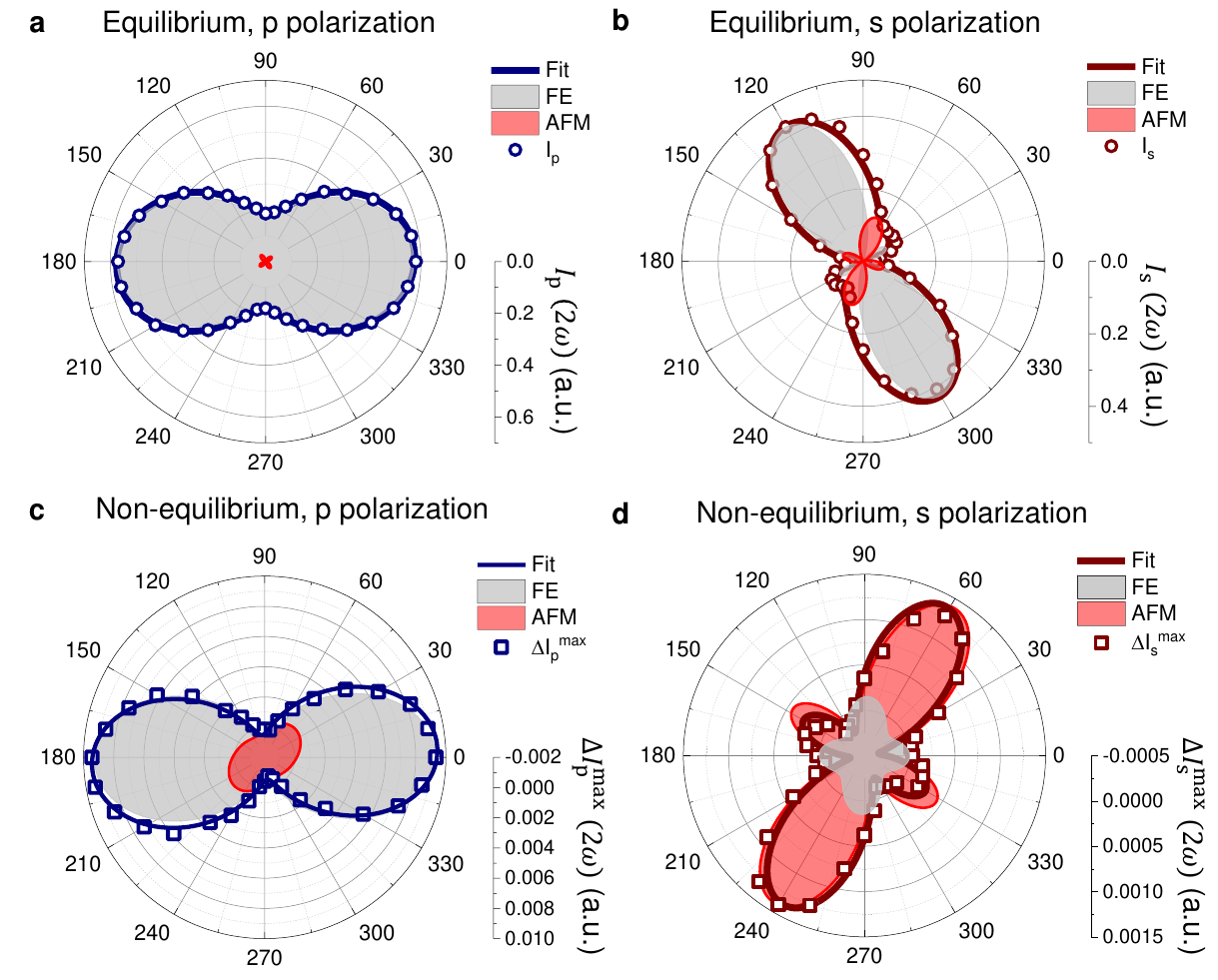}
\caption{\textbf{Static and time-resolved SHG polarimetry.} Polar plots of the equilibrium SHG intensity, $I$, are shown for \textbf{a,} $p$-polarization detection and \textbf{b,} $s$-polarization detection (open circles). Polar plots of the pump-induced change of the SHG intensity, $\Delta I$, are shown for \textbf{c,} $p$-polarization detection and \textbf{d,} $s$-polarization detection at the maximum of the pump-probe response (open squares). In all plots, solid lines are fits of the SHG intensities, composed of the ferroelectric contribution (grey areas) and the antiferromagnetic contribution (red areas). Both the ferroelectric and antiferromagnetic contributions show an enhancement following the resonant phonon excitation.
}
\label{fig:SHGandDSHG}
\end{figure*}

In Figs.~\ref{fig:SHGandDSHG}a and b, we show the probe-polarization dependence of the SHG signal in equilibrium. For the $p$-polarization detection, the SHG polar plot has two lobes at 0$^\circ$ and 180$^\circ$, respectively. This can be fitted using nearly exclusively the $\chi^{(i)}$ tensor components (ferroelectric contribution, grey area). For the $s$-polarization detection, the SHG signal has four lobes, with a clear asymmetry along the 60$^\circ$ and 240$^\circ$ directions. This asymmetry is due to the nonzero $\chi^{(c)}$ tensor components (antiferromagnetic contribution, red area) arising from the symmetry lowering in the single spin-cycloid domain. Similar features were observed in BiFeO$_3$ thin films, in which the orientation of the asymmetric SHG component was used to identify different antiferromagnetic domains \cite{Chauleau2017}. In Figs.~\ref{fig:SHGandDSHG}c and d, we show the SHG polar plots in the phonon-driven state at the maximum of the pump-probe response. For both the $p$- and $s$-polarization detections, $\Delta I^{\text{max}}$ is anisotropic and primarily positive with varying probe polarization. The maximum enhancement for both $\Delta I^{\text{max}}_\text{p}/I_\text{p}$ and $\Delta I^{\text{max}}_\text{s}/I_\text{s}$ is approximately (1.5$\pm$0.4)\%. Intriguingly, the dominant contributions to the pump-induced $p$- and $s$-signal changes are from different $\chi$ tensors. In the $p$-detection channel, the maximum enhancement in SHG intensity is along 0$^\circ$ and 180$^\circ$, given by the ferroelectric SHG tensor (grey area in Fig.~\ref{fig:SHGandDSHG}c). In contrast, in the $s$-detection channel, the maximum enhancement in SHG intensity is along 60$^\circ$ and 240$^\circ$, dominated by the antiferromagnetic SHG tensor (red area in Fig.~\ref{fig:SHGandDSHG}d). In order to shed light on the physical mechanisms that lead to this simultaneous enhancement of ferroelectric and antiferromagnetic SHG enhancement, we perform simulations of the nonlinear phonon dynamics that follow the coherent excitation of the high-frequency $A_1$ mode by the MIR pulse.

\subsection{Simulations of nonlinear phonon dynamics}\label{subsec3}

We evaluate the nonlinear vibrational response and the corresponding dynamics of the lattice polarization and the magnetic interactions in a semi-classical oscillator model, for which we compute the input parameters from first principles using density functional theory. The point-group symmetry of BiFeO$_3$ allows for quadratic-linear three-phonon couplings of the resonantly driven high-frequency $A_1$ mode (calculated eigenfrequency 15.3~THz) to the three low-frequency $A_1$ modes (calculated eigenfrequencies 8.8, 7.4, and 4.8~THz) present in the system, which we denote as $A^h_1$ and $A^l_1$ in the following. The coupling is described by an interaction potential of $V=cQ_{A^h_1}^2Q_{A^l_1}$, where $Q$ is the phonon amplitude and $c$ the coupling coefficient, and can be utilized for nonlinear phononic rectification: when the $A^h_1$ mode is coherently driven, it enacts a unidirectional force on the $A^l_1$ modes. This force induces a quasistatic displacement of the atoms along the eigenvectors of the $A^l_1$ modes that is proportional to the mean-squared amplitude of the $A^h_1$ mode, $\langle Q_{A^l_1}\rangle\propto\langle Q^2_{A^h_1}\rangle$. Furthermore, as the high-frequeny $A_1$ mode is driven with a large amplitude, it can experience self-rectification through cubic-order anharmonicity, described by the potential term $V=\tilde{c}Q_{A^h_1}^3$. Because all $A_1$ modes are infrared active, the rectification induces a quasistatic electric dipole moment and therefore an electric polarization in addition to the ferroelectric polarization. Details of the evolution of the phonon amplitudes according to the oscillator model and the density functional theory calculations of the coupling coefficients are provided in the Methods section and in the Supplementary Information Section S4.


\begin{figure}[t]
\centering
\includegraphics[width=1\textwidth]{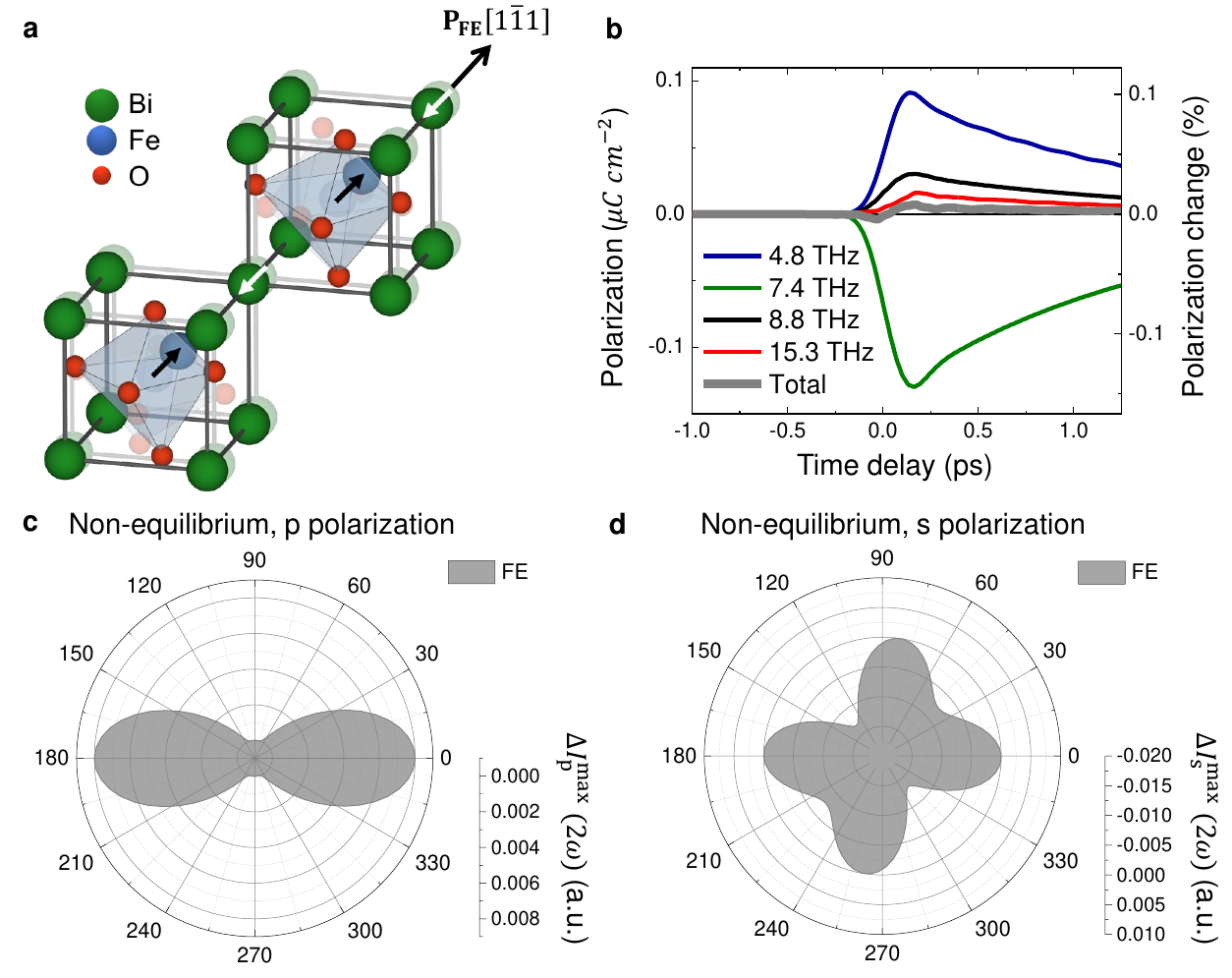}
\caption{\textbf{Simulations of nonlinear phonon dynamics and phonon-induced ferroelectric SHG.} 
\textbf{a,} Transient displacements of the ions along the rectified components of all four $A_1$ modes. Displacements are exaggerated for clarity. Besides the change in the oxygen octahedron, the two cations, bismuth and iron, move in opposite directions along the [1$\bar{1}$1] direction, marked by black and white arrows. 
\textbf{b,} Individual Polarization induced by the rectified components of the four $A_1$ modes. The left axis shows the polarization values, whereas the right axis shows percentile change of the electric polarization with respect to the equilibrium value ($90$~$\mu$C cm$^{-2}$). 
Calculated pump-induced changes in the SHG signal for the ferroelectric contribution are shown for $p$-polarization detection \textbf{(c)} and $s$-polarization detection \textbf{(d)} at the maximum response of the phonon amplitudes. The shape of the polar plot in $p$- and $s$-polarization reproduce the experimental data. While the computations accurately replicate the experimental magnitudes for $p$-polarization, there is a slight overestimation for $s$-polarization.}
\label{fig:dynamicscombined}
\end{figure}

In Fig.~\ref{fig:dynamicscombined}a, we show the displacements of the ions for the sum of the rectified components of all four $A_1$ modes. The cations, bismuth and iron, move in opposite directions along the [1$\bar{1}$1] direction. In Fig.~\ref{fig:dynamicscombined}b, we show the transient polarization induced by the rectified components of all four $A_1$ modes, as well as the relative change of lattice polarization compared to the equilibrium ferroelectric polarization of $90$~$\mu$C~cm$^{-2}$. The rectifications of the individual $A_1$ modes induce opposing polarizations due to the opposite signs of their mode effective charges that lead to only a small net change in polarization, which can be explained by the opposite net motion of cations shown in Fig.~\ref{fig:dynamicscombined}a.

In order to predict the pump-induced change in SHG intensity, we calculate and compare the ferroelectric contribution to the SHG tensor components of both the equilibrium and transiently distorted lattice structures at the maximum of the vibrational response (see Supplementary Information Section S4 for computational details). In Figure~\ref{fig:dynamicscombined}c and d, we show the polar plots of the change in SHG intensity, $\Delta I$, for the $p$- and $s$-detection channels that we obtain from the modified SHG tensor components due to nonlinear phononic rectification. The calculations well reproduce the experimental features (grey areas in Fig.~\ref{fig:SHGandDSHG}c and d), which confirms the ferroelectric contribution to the SHG enhancement of $\Delta I_p$, and supports the experimental finding that the enhancement of $\Delta I_s$ along 60$^\circ$ and 240$^\circ$ must stem from the antiferromagnetic $\chi^{(c)}$ tensor (red area in Fig.~\ref{fig:SHGandDSHG}d).

\begin{figure}[t]
\centering
\includegraphics[width=1\textwidth]{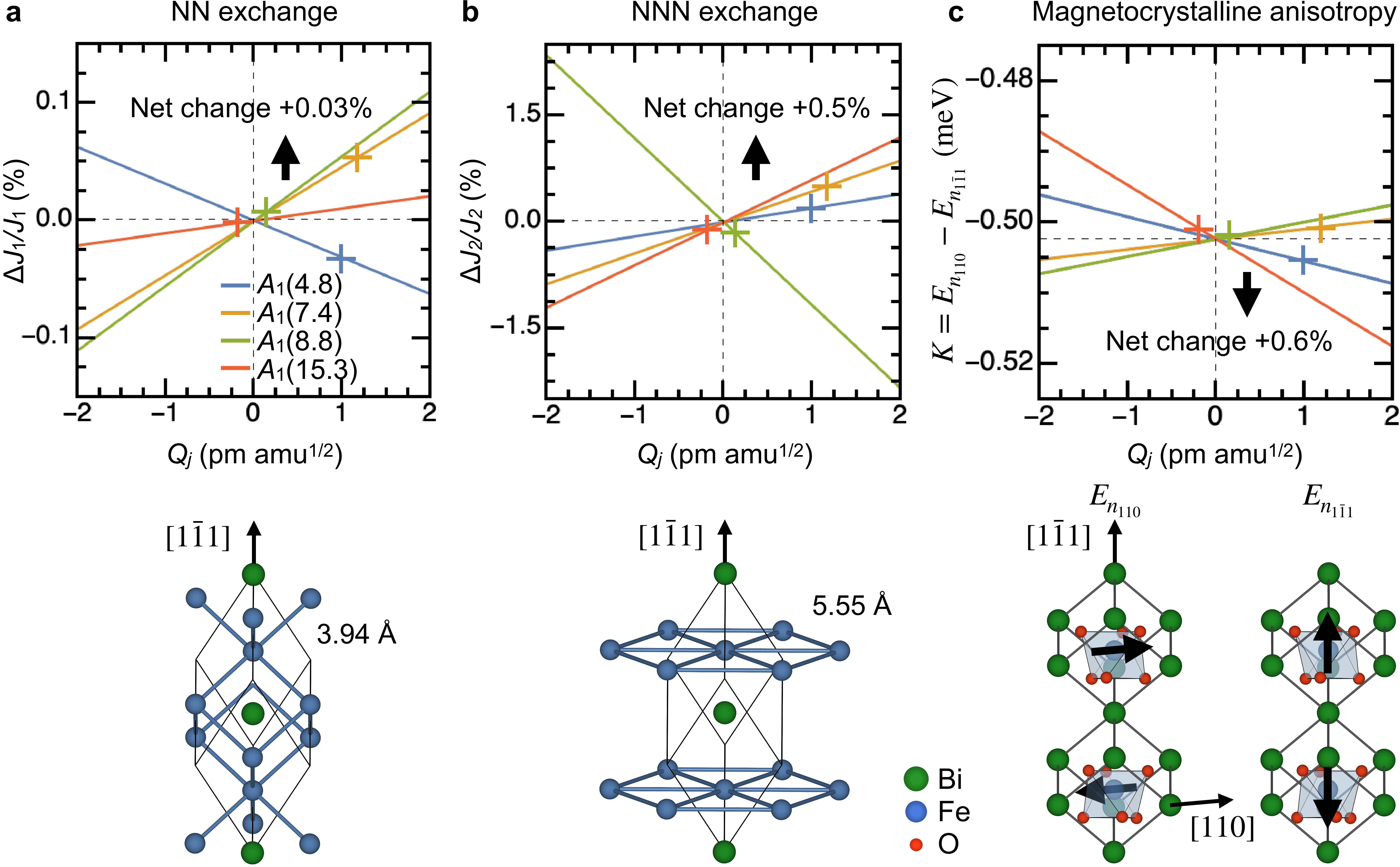}
\caption{\textbf{Simulations of phonon-dependent magnetic interactions.} We show the changes of the magnetic interactions induced by displacement of the atoms along the eigenvectors of the four $A_1$ modes for \textbf{a}, the nearest neighbor (NN) exchange interaction $J_1$, showing a net increase of $\Delta J_1/J_1=0.03\%$, \textbf{b}, the next nearest neighbor (NNN) exchange interaction $J_2$, showing a net increase of $\Delta J_2/J_2=0.5\%$, and \textbf{c}, the magnetocrystalline anisotropy $K$, computed as the energy difference between spin configurations oriented along [110] and [1$\bar{1}$1] directions, showing a net increase of $\Delta K/K=0.6\%$. Plus symbols mark the change of the magnetic interactions at the maximum rectified phonon amplitudes of the respective $A_1$ modes.
}
\label{fig:dynamicsmagneticorder}
\end{figure}

Finally, we investigate the impact of the driven structural state on the magnetic order by computing the changes in the exchange interactions and magnetocrystalline anisotropy of the iron spins that arise from the quasistatic displacements of the $A_1$ modes. (For details of the calculations, see Methods and Supplementary Information S4.4.) We show the changes of the magnetic interactions upon phonon displacement in Fig.~\ref{fig:dynamicsmagneticorder}. Intriguingly, we obtain enhancements of all magnetic interactions in the driven state, comparable in magnitude to the changes in electric polarization. Using the maximum rectified amplitudes for each of the $A_1$ modes (Supplementary Fig.~S4), we find that the nearest neighbor exchange $J_1$ (Fig.~\ref{fig:dynamicsmagneticorder}a) overall increases in magnitude by 0.03\%, the next nearest neighbor exchange $J_2$ (Fig.~\ref{fig:dynamicsmagneticorder}b) increases by 0.5\%{}, and the magnetocrystalline anisotropy (Fig.~\ref{fig:dynamicsmagneticorder}c)  increases by 0.6\%{}. This increase in both antiferromagnetic exchange interactions and the magnetocrystalline anisotropy indicates a more robust magnetic ordering in the driven state than in the ground state, which we suggest is at the heart of the enhancement of the antiferromagnetic SHG signal. Note however that we employed a simplified model, neglecting the spin spiral in the magnetic order of BiFeO$_3$. Consequently, these numbers should be used with caution and a more refined model would be necessary for a comprehensive quantitative discussion, as well as explicit calculations of the antiferromagnetic SHG tensors, which are beyond our current methodology.

\section{Discussion}\label{sec3}
We have demonstrated a simultaneous enhancement of ferroelectric and antiferromagnetic second-harmonic generation in BiFeO$_3$ by resonant excitation of a high-frequency fully-symmetric phonon mode. We stress that the pump-induced SHG intensity changes are not a multi-domain effect, as the single-crystal sample studied here is predominantly in a single ferroelectric and spin-cycloid domain. The SHG signals from minor domains are negligible and do not match the experimental data, see Supplementary Information Section S3. We can further exclude the contribution from MIR-induced SHG, which involves the third-order susceptibility tensor, $\chi^{(3)}$.The $\chi^{(3)}$ signal is spectrally filtered (see Methods) and the $\chi^{(3)}$ process is spatially separated from the $\chi^{(2)}$ process due to the 45$^\circ$ angle between the SHG probe and the MIR pump beams. In addition, the $\chi^{(3)}$ process would result in a quadratic pump fluence dependence for $\Delta P(2\omega)$, not the linear relation observed here (Fig.~\ref{fig:MIRandDeltaISHG}d). 

The enhancement of the ferroelectric contribution can be well reproduced by our calculations of the $\chi^{(i)}$ tensor-component modulations due to the transient lattice distortion induced by nonlinear phononic rectification. Intriguingly, the lattice polarizations of the $A_1$ phonon have opposite signs due to the counter-phase motion of the cations along the direction of ferroelectric polarization, $\mathbf{P}_\mathrm{FE}$, and nearly cancel each other out. In turn, they enhance the SHG tensor components directly, therefore leading to a larger $\mathbf{P}(2\omega)$ in comparison to the equilibrium state that our measurements pick up. While our theory captures the ferroelectric SHG features well, a direct calculation of the antiferromagnetic SHG tensor is beyond the scope of this work. Our calculations indicate however that both the exchange interaction and magnetocrystalline anisotropy, which are responsible for the magnetic ordering, are enhanced by the rectified displacements of the $A_1$ modes. This mechanism of magnetic-interaction control through phonons has been shown for nonpolar magnets before \cite{Fechner2018,Khalsa2018,Rodriguez-Vega2020,Padmanabhan2022}. As a result, this modification leads to a more robust magnetic ordering and likely to changes of the antiferromagnetic $\chi^{(c)}$ tensor components, similar to the ferroelectric case. The fully-symmetric $A_1$ modes by definition preserve the symmetry of the system and while they leave the relative distances between the iron ions mostly unchanged, they modify the relative distances between the iron and oxygen ions. This could further affect the Dzyaloshinskii-Moriya interaction that is responsible for the noncollinear arrangement of spins in BiFeO$_3$.

Finally, we briefly touch on other possible mechanisms of phonon-induced magnetization control, which could be at play in addition to the enhancement of magnetic interactions. One possibility concerns phonon-induced symmetry breaking of the orbital configuration, which has recently been shown to stabilize magnetization above the ordering temperature in a transition-metal ferromagnet \cite{Disa2023}. Due to the fully-symmetric nature of the $A_1$ modes, we can likely rule out this mechanism in our case. Another possibility concerns coherent magnon excitation through coupling to the phonons, which has in the past been described as ionic Raman scattering by magnons \cite{nova:2017,juraschek2:2017,Juraschek2020_3}. While ionic Raman scattering of the $A_1$ modes by the magnons might be possible in BiFeO$_3$, only circularly polarized phonons would lead to a unidirectional force on the spins \cite{Dominik2022}. All $A_1$ modes involved in the dynamics are linearly polarized, and we therefore also rule out this mechanism for our case. Furthermore, the spectrum of our MIR pulse is far away from the electromagnon resonances in BiFeO$_3$ \cite{DeSousa2008,Cazayous2008}, which allows us to exclude a coupling to the spins through the dipole moment of the spin cycloid. With the data and calculations given thus far, we therefore find it most likely that the modifications of the exchange interactions and magnetocrystalline anisotropy lead to a stabilization of the antiferromagnetic order and to a change of the $\chi^{(c)}$ components.

\section{Conclusion}\label{sec4}

To conclude, our results show that ultrashort pulses in the THz and MIR spectral range can be used to address multiple ferroic order parameters at once, utilizing the coupling to nonlinearly driven phonon modes. We have demonstrated simultaneous ferroelectric and antiferromagnetic SHG enhancement for the case of the prototypical multiferroic material BiFeO$_3$, but we expect that this approach will be applicable generically to materials exhibiting magnetoelectric coupling. The ferroic orders in BiFeO$_3$ form independently below the respective Curie and N\'{e}el temperatures at 1100 and 640~K. In so-called type-II multiferroics, in contrast, the ferroic orders are interdependent and emerge together, but these materials often suffer from very low multiferroic ordering temperatures \cite{Cheong2007}. An intriguing question therefore arises, whether both ferroic orders can be stabilized simultaneously above the multiferroic ordering temperature through coherent phonon excitation. This problem can be regarded in the same broad context as the light-induced stabilization of superconductivity, ferroelectricity, and magnetism demonstrated in recent experiments \cite{mankowsky:2014,Mitrano2016,Nova2019,Li2019,Disa2023}.

\setcounter{figure}{0}
\renewcommand{\thefigure}{M\arabic{figure}}

\section{Methods}\label{sec5}

\subsection{Crystal growth}
The BiFeO$_3$ single crystal was grown using a traveling solvent technique in a Laser Floating Zone (LFZ) furnace \cite{Ito2011}. Bi$_2$O$_3$ (99.975\%, Alfa Aesar) and Fe$_2$O$_3$ (99.995\%, Alfa Aesar) powder are mixed in molar ratio 1 : 1 and 4 : 1 for the feed rod and the traveling solvent, respectively. The feed rod material was sintered at 800$^{\circ}$C for 20 hours with one intermediated grinding, and the solvent was sintered at 700$^{\circ}$C for 5 hours. A 0.15 g solvent pellet was placed on top of the seed rod before starting the floating zone growth. The growth was performed in air, and the growth rate was 1 mm~h$^{-1}$. The as-grown crystal is oriented by a Laue x-ray camera. The ferroelectric domain structure has been characterized by optical microscope, circular interference differential contrast and piezoresponse force microscopy imaging (Supplementary Information S3). 
\subsection{MIR pump-SHG probe setup}\label{subsec5}
\begin{figure}
\centering
\includegraphics[width=0.7\textwidth]{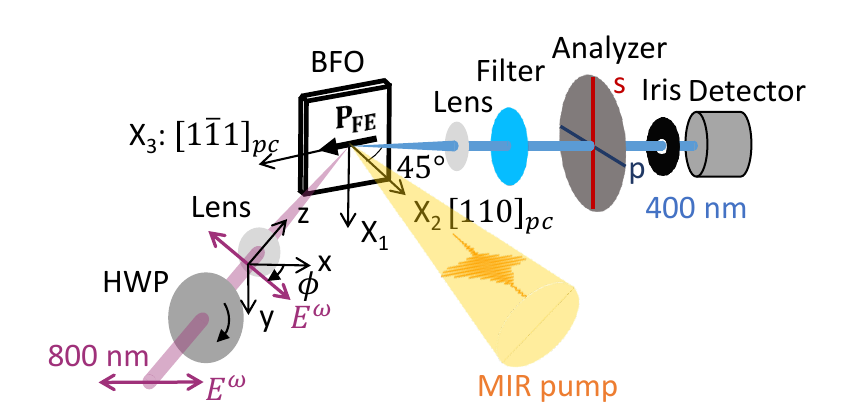}
\caption{Schematic of the MIR pump-SHG probe experiment, with the orientations of the $(x, y, z)$ laboratory frame and the $(X_1, X_2, X_3)$ crystal frame indicated. A half-wave plate (HWP) rotates the polarization of the 800~nm pulses. A dichroic mirror (not shown) reflects the second-harmonic generation signal and removes the residual 800~nm signal. A band-pass filter (10~nm bandwidth) removes signals from higher-order optical processes. An analyzer selects either the vertically polarized ($s$) or horizontally polarized ($p$) SHG signal.} \label{fig:SHGSetup}
\end{figure}

The fundamental beam for SHG is from a Ti:sapphire laser with a repetition rate of 1~kHz, energy of 30~nJ, and a duration of 40~fs. A schematic representation of the setup is shown in Fig.~\ref{fig:SHGSetup}. A half-wave plate (HWP) was used to rotate the polarization of the fundamental beam. The beam was focused SHG signal was spectrally filtered by a band-pass filter, spatially filtered by an iris, and then detected by a photo-multiplier tube. A Glan-Taylor prism was used to select the desired polarization of the second-harmonic signal. All optical measurements were performed at room temperature, with a 45$^\circ$ incidence in reflection geometry.

MIR pump pulses with tunable wavelengths between 13 and 18.5~micron, and a duration of 250~fs were generated by optical parametric amplification and subsequent difference frequency generation in a GaSe crystal. The MIR pulses were focused on to the sample at normal incidence, with a diameter of 220~micron and a maximum excitation fluence of 12~mJ~cm$^{-2}$.

\subsection{Fits of the SHG signal}
At room temperature, BiFeO$_3$ crystallizes in rhombohedral symmetry with point group ${3m}$ and space group $R3c$. We set $X_3$ as the ferroelectric axis and $X_2$ as the normal to the mirror plane of the trigonal structure (Fig.~\ref{fig:SHGSetup}). In a single antiferromagnetic domain, the long-range cycloidal modulation of the spin breaks the three-fold rotation symmetry around $\mathbf{P}_\mathrm{FE}$, reducing the symmetry of the material to monoclinic with point group $m$. Details of the SHG tensors can be found in Supplementary Information Section S2. The fit reveals that the cycloidal spin modulation direction is along $X_2$.  

\subsection{Theory of nonlinear phonon dynamics}\label{subsec6}
We use a phenomenological oscillator model to evaluate the nonlinear vibrational response of the $A_1$ modes and the corresponding lattice-polarization dynamics following the excitation by an ultrashort MIR pulse, for which we compute all parameters from first principles using density functional theory. Specifically, we solve the equations of motion
\begin{equation}\label{eq:eom}
\ddot{Q}_i + \kappa_i \dot{Q}_i + \partial_{Q_i} V = 0,
\end{equation}
where $Q_i$ are the normal mode coordinates (amplitudes) and $\kappa_i$ the phenomenological linewidths of the $A_1$ modes in BiFeO$_3$, $i\in\{A_1\}$. $V = V_0 + V_\mathrm{int}$ is the potential energy, consisting of an equilibrium potential, $V_0$, and an interaction potential, $V_\mathrm{int}$. We expand $V_0$ up to fourth order in anharmonic and nonlinear coupling terms,
\begin{equation}\label{eq:potential}
V_0 = \sum\limits_{i=\{A_1\}} \frac{\Omega_i^2}{2}Q_i^2 + \sum\limits_{i,j,k=\{A_1\}} c_{ijk} Q_i Q_j Q_k + \sum\limits_{i,j,k,l=\{A_1\}} d_{ijkl} Q_i Q_j Q_k Q_l,
\end{equation}
where $\Omega_i$ is the phonon eigenfrequency, and $c_{ijk}$ and $d_{ijkl}$ are the anharmonic and nonlinear coupling coefficients. Because the $A_1$ modes are fully symmetric, all permutations of modes are allowed by symmetry in the anharmonic terms. We take into account all single-mode anharmonicities ($c_{iii}$, $d_{iiii}$), but only nonlinear couplings of the high-frequency $A_1$(15.3 THz) mode, which is resonantly driven, to the other $A_1$ modes of the system. See Supplementary Information Section S4 for the density functional theory calculations of the coupling coefficients.

The interaction potential, $V_\mathrm{int}$, contains the coupling of the electric dipole moment of the infrared-active phonon $A_1$ modes, $p_i$, to the electric field component of the laser pulse, $E(t)$, and can be written as
\begin{equation}\label{eq:interaction}
V_\mathrm{int} = \sum\limits_{i=\{A_1\}} p_i E(t) = \sum\limits_{i=\{A_1\}} Z_i Q_i E(t),
\end{equation}
where the linear polarization of the laser pulse is assumed to be aligned along the ferroelectric polarization direction of the crystal. $Q_i$ is the phonon amplitude as evaluated in Eq.~(\ref{eq:eom}) and $Z_i$ is the mode effective charge given by $Z_i = \sum_{n} Z^*_n q_{n,i}/\sqrt{M_n}$. $Z^*_n$ is the Born effective charge tensor, $q_{n,i}$ the phonon eigenvector, and $M_n$ the ionic mass of ion $n$, and the index $n$ runs over all ions in the unit cell. We model the electric field component of the laser pulse as $E(t) = E_0 \exp[-t^2/(2(\tau_0/\sqrt{8\ln2})^2)] \cos(\omega_0 t)$ and set its parameters to the experimental values: a full width at half maximum (FWHM) duration of 250~fs, a central frequency of 16.9~THz, and a fluence of 12~mJ~cm$^{-2}$ (peak electric field $E_0=5.3$~MV~cm$^{-1}$). Because $\omega_0$ is far off resonance from the three low-frequency $A_1$ modes, only the contribution of the $A_1$(15.3 THz) mode plays a role in Eq.~(\ref{eq:interaction}). Finally, the time-dependent electric polarization, $P(t)$, induced by the displacements of the ions along the coordinates of the infrared-active $A_1$ modes, is given by 
\begin{equation}\label{eq:polarization}
P(t) = \sum\limits_{i=\{A_1\}} p_i(t)/V_c,
\end{equation}
where $V_c$ is the volume of the unit cell. 

We next compute the phonon-induced modulation of the SHG tensor components. Due to the lack of inversion symmetry in the crystal, any infrared-active mode can linearly modify the components of $\chi^{(2)}$ as
\begin{equation}
\chi_\text{pumped}^{(2)}(t) = \chi_\text{equilibrium}^{(2)} + \sum_{j = \{A_1\}} \frac{\partial\chi_j^{(2)}}{\partial Q_j} Q_j(t).
\end{equation}
Here, $\chi_\text{equilibrium}^{(2)}$ is the equilibrium second-order susceptibility and $\partial_{Q_j}\chi_j^{(2)}$ is the phonon-induced change. We compute both the equilibrium and nonequilibrium tensor components for the ferroelectric contribution, $\chi^{(i)}$, from first principles, see Supplementary Information Section S4.

We finally compute the time-dependent modifications of the exchange interactions, $J_1$ and $J_2$, as well as the magnetocrystalline anisotropy, $K$, as a result of the rectified $A_1$ mode displacements
\begin{align}
    J_{1/2}(t) & = J_{1/2}^\text{equilibrium} +  \sum_{j = \{A_1\}} \frac{\partial J_{1/2}}{\partial Q_j}Q_j(t), \\
    (E_{n_{110}}-E_{n_{1\bar{1}1}})(t)\equiv K(t) & = K^\text{equilibrium} + \sum_{j = \{A_1\}} \frac{\partial K}{\partial Q_j}Q_j(t).
\end{align}
We calculate $J_{1/2}^\text{equilibrium}$, $K^\text{equilibrium}$, $\partial_{Q_j}J_{1/2}$, and $\partial_{Q_j}K$ from first principles, see Supplementary Information Section S4.4.

\section*{Supplementary Information}
Details of the SHG analysis and density functional theory calculations can be found in the Supplementary Information. 
\section*{Acknowledgments}
We acknowledge the helpful discussions with Liuyan Zhao, Ankit Disa, and Rui Zu.
D.A.B.L. and W.H acknowledge support from the U.S. Department of Energy, Office of Science, Office of Basic Energy Sciences Early Career Research Program under Award Number DE-SC-0021305. The work at Rutgers was supported by the W.M. Keck Foundation. D.M.J. was supported by Tel Aviv University.
\section*{Author contributions}
D.A.B.L. and D.M.J. contributed equally to this work. W.H. conceived the project, designed and performed the pump-probe measurements, and analyzed the data. D.A.B.L. performed data fitting and modeling. D.M.J. performed the density functional theory calculations and simulations of the nonlinear phonon dynamics. M.F. performed the density functional theory calculations of the SHG tensor components. X.X. and S-W.C. grew the sample. W.H. and D.M.J. wrote the paper with input from all other authors.

\section*{Competing interests}
All authors declare no competing interests.

\section*{Data Availability}
The authors declare that all data supporting the findings of this study are available within the article and its Supplementary Information. 




\end{document}